\documentclass[conference]{IEEEtran}
\IEEEoverridecommandlockouts
\usepackage{cite}
\usepackage{amsmath,amssymb,amsfonts}
\usepackage{algorithmic}
\usepackage{graphicx}
\usepackage{textcomp}
\usepackage{xcolor}
\usepackage{multirow}

\usepackage{hyperref}
\usepackage{url}

\usepackage{verbatim}

\usepackage{url}
\usepackage{tabularx}  

\usepackage{color}
\usepackage{tikz}
\usetikzlibrary{patterns}
\usetikzlibrary{calc}

\newcommand{\convexpath}[2]{
  [   
  create hullcoords/.code={
    \global\edef\namelist{#1}
    \foreach [count=\counter] \nodename in \namelist {
      \global\edef\numberofnodes{\counter}
      \coordinate (hullcoord\counter) at (\nodename);
    }
    \coordinate (hullcoord0) at (hullcoord\numberofnodes);
    \pgfmathtruncatemacro\lastnumber{\numberofnodes+1}
    \coordinate (hullcoord\lastnumber) at (hullcoord1);
  },
  create hullcoords
  ]
  ($(hullcoord1)!#2!-90:(hullcoord0)$)
  \foreach [
  evaluate=\currentnode as \previousnode using \currentnode-1,
  evaluate=\currentnode as \nextnode using \currentnode+1
  ] \currentnode in {1,...,\numberofnodes} {
    let \p1 = ($(hullcoord\currentnode) - (hullcoord\previousnode)$),
    \n1 = {atan2(\y1,\x1) + 90},
    \p2 = ($(hullcoord\nextnode) - (hullcoord\currentnode)$),
    \n2 = {atan2(\y2,\x2) + 90},
    \n{delta} = {Mod(\n2-\n1,360) - 360}
    in 
    {arc [start angle=\n1, delta angle=\n{delta}, radius=#2]}
    -- ($(hullcoord\nextnode)!#2!-90:(hullcoord\currentnode)$) 
  }
}

\def\BibTeX{{\rm B\kern-.05em{\sc i\kern-.025em b}\kern-.08em
    T\kern-.1667em\lower.7ex\hbox{E}\kern-.125emX}}
\begin{document}

\title{Automatic coherence-driven inference on arguments \\
}

\author{\IEEEauthorblockN{Steve Huntsman}
\IEEEauthorblockA{sch213@nyu.edu}}

\maketitle

\begin{abstract}
Inconsistencies are ubiquitous in law, administration, and jurisprudence. Though a cure is too much to hope for, we propose a technological remedy. Large language models (LLMs) can accurately extract propositions from arguments and compile them into natural data structures that enable \emph{coherence-driven inference} (CDI) via combinatorial optimization. This neurosymbolic architecture naturally separates concerns and enables meaningful judgments about the coherence of arguments that can inform legislative and policy analysis and legal reasoning.
\end{abstract}


\section{\label{sec:introduction}Introduction}

Imagine that public officials could be nudged \cite{thaler2021nudge} to develop more coherent bodies of policy, law, and jurisprudence \cite{cyrul2013consistency,savelka2013coherence,kahneman2021noise}. Imagine that individuals could be nudged towards epistemic virtue in reasoning about the coherence of their observations and beliefs. 
Imagine an approach to artificial intelligence that was capable of doing these things in a way that could be trusted and verified.

\begin{figure}[htbp]
  \centering
  \fbox{\includegraphics[trim = 83mm 100mm 75mm 115mm, clip, width=.16\columnwidth,keepaspectratio]{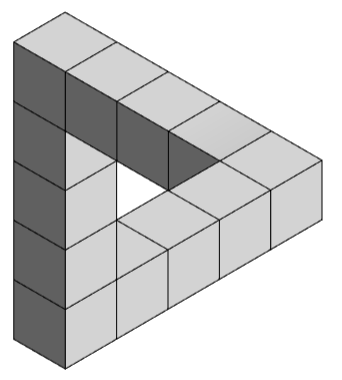}}
  \fbox{\includegraphics[trim = 83mm 100mm 75mm 115mm, clip, width=.16\columnwidth,keepaspectratio]{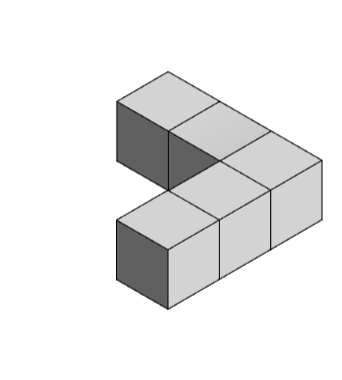}}
  \fbox{\includegraphics[trim = 83mm 100mm 75mm 115mm, clip, width=.16\columnwidth,keepaspectratio]{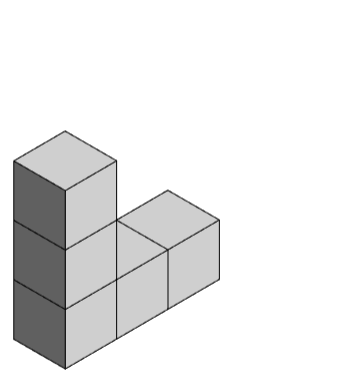}}
  \fbox{\includegraphics[trim = 83mm 100mm 75mm 115mm, clip, width=.16\columnwidth,keepaspectratio]{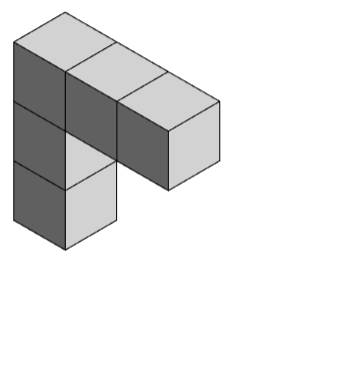}}
  \fbox{\includegraphics[trim = 83mm 100mm 75mm 115mm, clip, width=.16\columnwidth,keepaspectratio]{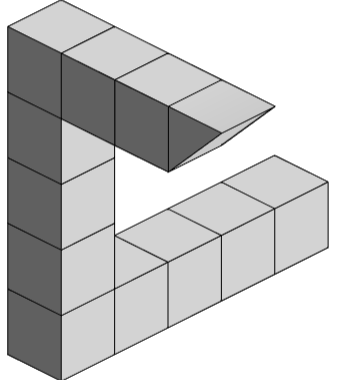}}
  \caption{The Penrose triangle in the left panel cannot be realized by consistently gluing together local data suggested by the middle three panels, i.e., the cubes at the ends of the three L shapes. The underlying mathematics informs the solution of constraint satisfaction problems \cite{rosiak2022sheaf,bach1999sheaf}, and motivated a rediscovery and generalization of CDI \cite{huntsman2024prospects}.
  }
  \label{fig:penrose}
\end{figure}

An approximation of this sort of Utopia is actually plausible, as we indicate below. The underlying idea is a neurosymbolic architecture \cite{sarker2022neuro,marra2024statistical} for artificial intelligence that is fundamentally rooted in resolving incoherence and ambiguity. This architecture cleanly separates concerns between local ``fast/system 1'' inference \cite{kahneman2011thinking} via large language models (LLMs) and global ``slow/system 2'' inference via combinatorial optimzation, called \emph{coherence-driven inference} (CDI) \cite{thagard1989explanatory,thagard1998coherence,thagard2002coherence,blokpoel2025theoretical}. In this architecture, LLMs compile local quantitative judgments about information consistency to a global data structure with high fidelity, so that CDI can proceed automatically \cite{huntsman2025neurosymbolic}. Figure \ref{fig:penrose} provides a cartoon of global incoherence: rejecting some local data (such as the second panel in the figure) can resolve incoherence. 

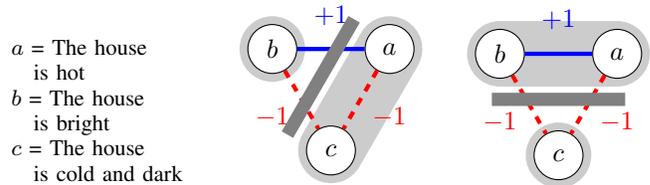
\begin{figure}[htbp]
    \centering   
        \resizebox{.3\columnwidth}{!}{%
            \begin{tikzpicture}[
              xscale=4, 
              yscale=4,
            ]
            \node[align=left,font=\footnotesize] (a) at (0,0) {
                $a$ = The house \\ \quad is hot \\
                $b$ = The house \\ \quad is bright \\
                $c$ = The house \\ \quad is cold and dark
                 };
            \end{tikzpicture}
        }    
        \quad
        \begin{tikzpicture}[scale=.90]
            \coordinate (A) at (30:1);
            \coordinate (B) at (150:1);
            \coordinate (C) at (270:1);
            \draw[thick,black!0,fill=black,opacity=0.2] \convexpath{A,C}{5mm};
            \draw[thick,black!0,fill=black,opacity=0.2](B) circle (5mm) node {};
            \node (q1) at (90:1) {{\color{blue}$+1$}};
            \node (q2) at (210:1) {{\color{red}$-1$}};
            \node (q3) at (330:1) {{\color{red}$-1$}};
            \node [draw,circle,fill=white,minimum size=6.5mm] (a) at (A) {$a$};
            \node [draw,circle,fill=white,minimum size=6.5mm] (b) at (B) {$b$};
            \node [draw,circle,fill=white,minimum size=6.5mm] (c) at (C) {$c$};
            \foreach \from/\to in {
                a/b}
                \draw[ultra thick, blue] (\from) to (\to);
            \foreach \from/\to in {
                a/c,b/c}
                \draw[ultra thick, red, dashed] (\from) to (\to);
            \coordinate (cut1) at (70:1);
            \coordinate (cut2) at (230:1);
            \draw[line width=5pt, color=gray] (cut1) to (cut2);
        \end{tikzpicture}
        \quad
        \begin{tikzpicture}[scale=.90]
            \coordinate (A) at (30:1);
            \coordinate (B) at (150:1);
            \coordinate (C) at (270:1);
            \draw[thick,black!0,fill=black,opacity=0.2] \convexpath{A,B}{5mm};
            \draw[thick,black!0,fill=black,opacity=0.2](C) circle (5mm) node {};
            \node (q1) at (90:1) {{\color{blue}$+1$}};
            \node (q2) at (210:1) {{\color{red}$-1$}};
            \node (q3) at (330:1) {{\color{red}$-1$}};
            \node [draw,circle,fill=white,minimum size=6.5mm] (a) at (A) {$a$};
            \node [draw,circle,fill=white,minimum size=6.5mm] (b) at (B) {$b$};
            \node [draw,circle,fill=white,minimum size=6.5mm] (c) at (C) {$c$};
            \foreach \from/\to in {
                a/b}
                \draw[ultra thick, blue] (\from) to (\to);
            \foreach \from/\to in {
                a/c,b/c}
                \draw[ultra thick, red, dashed] (\from) to (\to);
            \coordinate (cut1) at (-10:1);
            \coordinate (cut2) at (190:1);
            \draw[line width=5pt, color=gray] (cut1) to (cut2);
        \end{tikzpicture}
    \caption{
    (From \cite{huntsman2025neurosymbolic}.) 
    Left: three propositions labeled $a$, $b$, and $c$. Center: the corresponding coherence graph $G$. {\color{blue}Consistent (solid blue)} and {\color{red}inconsistent (red dashed)} proposition pairs/edges respectively inherit {\color{blue}positive} and {\color{red}negative} weights. We show the partition/cut $\{\{a, c\},\{b\}\}$ whose coherence is $-(A_{ab} + A_{bc}) = -(({\color{blue}+1}) + ({\color{red}-1})) = 0$, where $A$ is the weighted adjacency matrix of $G$. Right: the bipartition $\{\{a, b\},\{c\}\}$ whose coherence is $-(A_{ac} + A_{bc}) = -(({\color{red}-1}) + ({\color{red}-1})) = 2$. CDI amounts to partitioning data into an accepted set $U$ and the complementary rejected set so as to maximize the coherence objective \eqref{eq:coherence}. The bipartition on the right is optimal: it cuts across inconsistencies.
    }
    \label{fig:cdi}
\end{figure}

In CDI, an individual datum (here, a proposition or claim, e.g., in a regulatory document \cite{kumar2024nlp}) is represented by a vertex in a graph, as in Figure \ref{fig:cdi}. A \emph{coherence graph} has edges and weights in $[-1,1]$ that respectively represent relevance and consistency relations between data. The objective of CDI is maximizing \emph{coherence}, defined on partitions of the vertices of a coherence graph into accepted and rejected sets by the negative sum of weights for edges that have one vertex in each part. The optimal objective corresponds to a maximum cut in the weighted coherence graph. 

That is, for a weighted coherence graph $G = (V,E,w)$ with edge weights $w: E \rightarrow [-1,1]$, the coherence of a bipartition $\{U,V-U\}$ of graph vertices is
\begin{equation}
    \label{eq:coherence}
    \text{coherence}(\{U,V-U\}) := -\sum_{u \in U, v \not \in U} A_{uv},
\end{equation} 
where $A$ is the weighted adjacency matrix of $G$. 
\footnote{
Some other expressions of coherence in the literature are superficially different but completely equivalent up to irrelevant constants.
}
While this is a very hard computational problem (specifically, the $\mathbf{APX}$-complete MAX-CUT problem \cite{khot2007optimal,moore2011nature,gartner2012approximation,lee2021classifying}), it can be solved exactly at useful scales and it can be solved approximately at larger scales. 
\footnote{The examples in this paper are already at scales similar to those in the existing literature on CDI. Directly scaling to larger inferences is probably not particularly useful compared to structuring inference hierarchically and/or sequentially in closer correspondence to human cognition. Naive scaling to larger coherence graphs in the hope of more stronger cognitive abilities is of questionable utility in light of a fixed-parameter tractable cognition hypothesis \cite{van2008tractable,van2019cognition}, though efficient approximation algorithms also exist and coherence graphs themselves can be approximated by so-called cut sparsifiers that inform the approach \cite{huntsman2025neurosymbolic}.}
There is always at least one assignment of truth values that is optimal, though near-optimal assignments are of interest in the spirit of \emph{Rashomon} \cite{semenova2022existence}. It is possible to combine the most optimal assignments into a distribution over truth values in a principled way by using a Gibbs distribution \cite{huntsman2025cyber}.

\begin{figure}[htbp]
    \centering
    \includegraphics[width=1\linewidth, trim={75 25 1075 25mm}, clip]{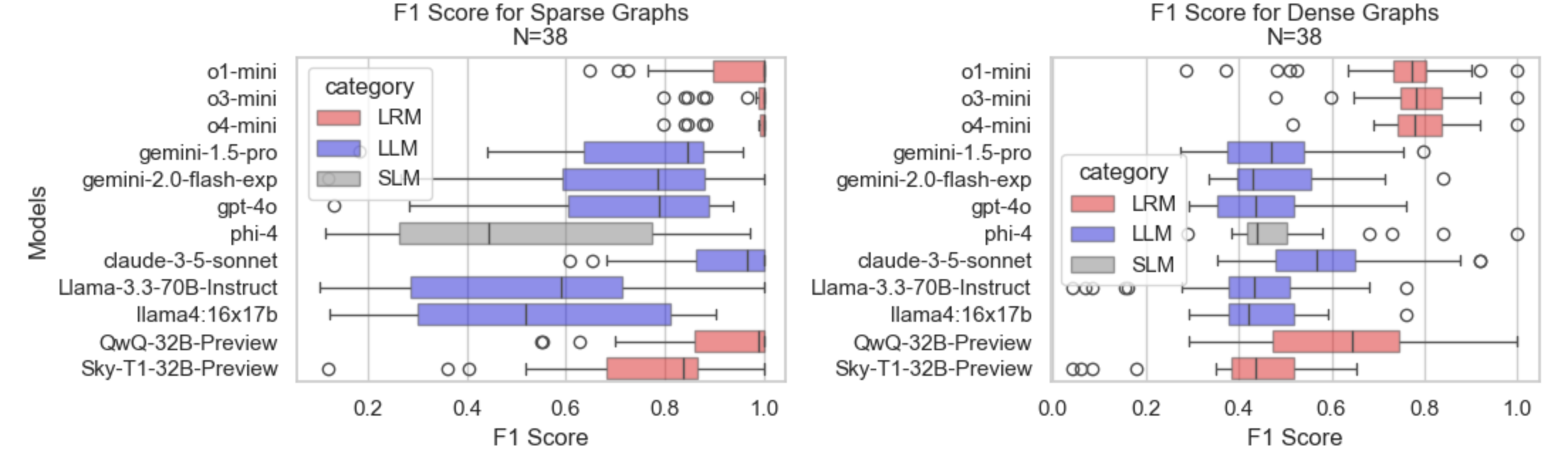}
        \caption{
        (Taken from \cite{huntsman2025neurosymbolic}, for which see details: 
        LRM and SLM respectively indicate large reasoning and small language models.) Some models have high micro $F_1$ scores on sparse coherence graph reconstruction tasks.}
    \label{fig:cohere_graphs}
\end{figure}

An algorithmic benchmark demonstrates that some LLMs (e.g., o1/3/4) can reproducibly (re)construct coherence graphs with high fidelity from propositions using a single prompt \cite{huntsman2025neurosymbolic}. This suggests a completely automatic approach to CDI since computing coherence is mechanical once a coherence graph is constructed. 
\footnote{
Inferring a coherence graph from a set of propositions is similar to classical natural language inference or textual entailment \cite{korman2018defining}, which involves evaluating the logical relationship between two sentences. With this in mind, \cite{huntsman2025neurosymbolic} briefly mentions inconclusive preliminary experiments that tried to identify either relevance or consistency directly from white box LLM embeddings or from various simple quantifications of inter-token attention. 
}
Figure \ref{fig:cohere_graphs} shows performance on this benchmark.

In \S \ref{sec:oval}-\ref{sec:brown}, we show examples of how automatically constructed median coherence graphs can inform analysis of real(istic) arguments. The idea is to extract the most important propositions in a transcript or report using a LLM, then to compile a coherence graph over these propositions, again using a LLM (in fact, doing this several times and taking a median of the vectorized adjacency matrix for robustness and stability), and finally to perform CDI on the result. 

Efficiently, reliably, and reproducibly identifying incoherence in arguments could improve legislation, administration, and jurisprudence: it could also strengthen civil society. CDI also has obvious applications to detecting, understanding, and combating threats to democracy such as conspiracy theories, propaganda, disinformation, etc., as well as other applications to machine cognition, but our focus in this paper is on a small set of illustrative argument examples.

Towards that end, the paper is organized as follows. In \S \ref{sec:local} we illustrate how even deprecated LLMs are capable of good fine-grained judgments of the consistency of propositions. (The ability of some LLMs to compile propositions in rigidly templated natural language into a coherence graph has also been demonstrated in \cite{huntsman2025neurosymbolic}, but space and complexity preclude a detailed overview of that result here.) In \S \ref{sec:examples} we present examples of four famous arguments and illustrate how CDI can usefully operate over key propositions extracted by LLMs. In \S \ref{sec:law} we discuss (in)coherence in law before concluding in \S \ref{sec:conclusion}. Appendices illustrate convergence properties of our constructions and the prompt used for compiling coherence graphs.

The code that we used to produce the examples in \S \ref{sec:examples} is available at \cite{huntsman2025argument}.

\section{\label{sec:local}LLMs can accurately determine local consistency of propositions}

\begin{figure}[htbp]
  \centering
  \includegraphics[trim = 15mm 50mm 20mm 60mm, clip, width=\columnwidth,keepaspectratio]{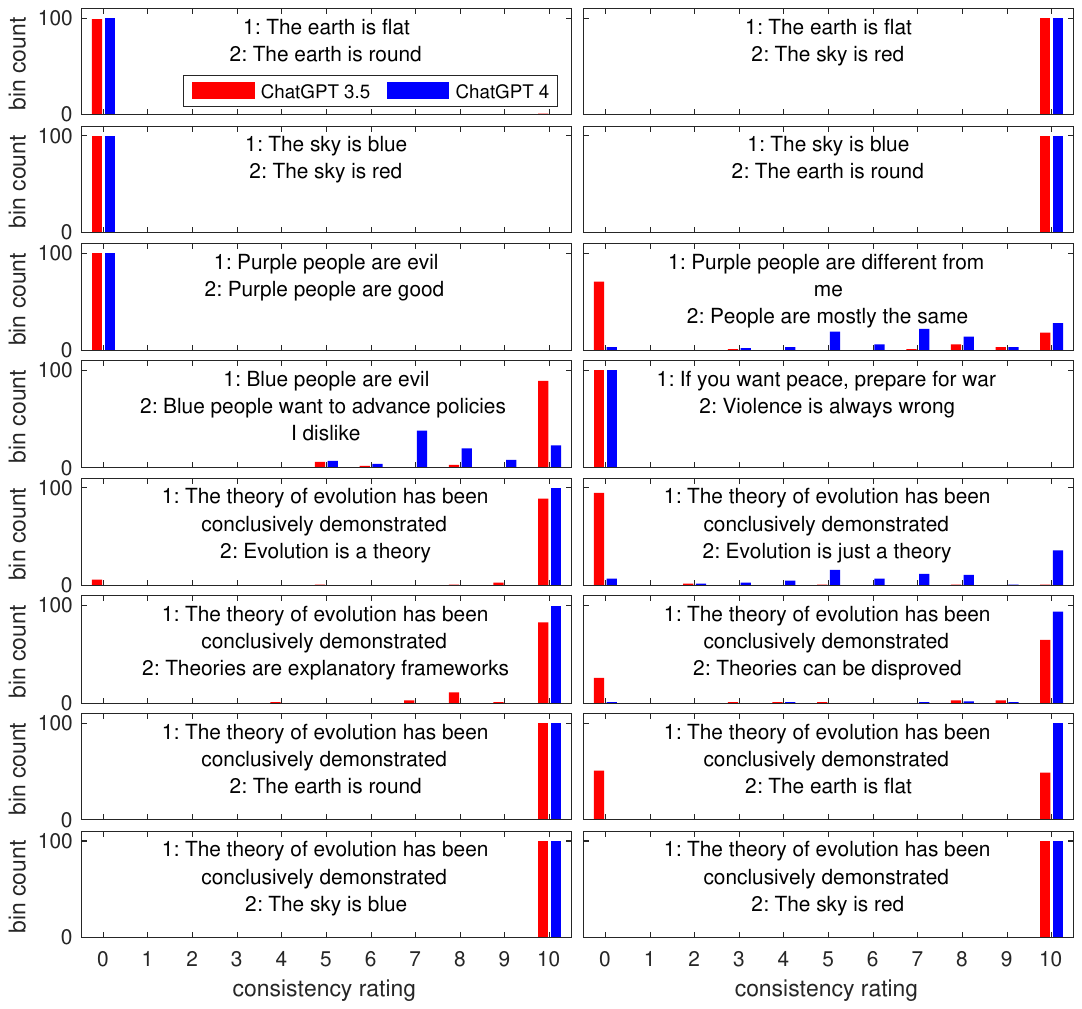}
  \caption{Histograms of $N = 100$ numerical consistency ratings produced by two versions of ChatGPT for the two propositions indicated in each panel.
  }
  \label{fig:consistency}
\end{figure}

Figure \ref{fig:consistency} shows the results of an experiment from the preprint \cite{huntsman2024prospects} 
in which ChatGPT 3.5-turbo and ChatGPT 4 rated the logical consistency of various pairs of propositions from 0 (inconsistent) to 10 (consistent). 

The results are reasonable and the underlying chain of thought appears to be defensible even if sometimes flawed \cite{huntsman2024prospects}.
The outputs from ChatGPT 4 appeared to be more sophisticated and reliable, as indicated by the increased variance for more ambiguously related pairs of propositions. We have not observed any hallucinations in these settings, and there is a simple reason: LLMs are just analyzing the information within a prompt, \emph{interpolating} between two propositions rather than \emph{extrapolating} as in many applications of generative artificial intelligence. More recent LLMs are likely to perform at least as well as the ones discussed in this section.

\begin{figure}[htbp]
  \centering
  \includegraphics[trim = 30mm 110mm 30mm 110mm, clip, width=\columnwidth,keepaspectratio]{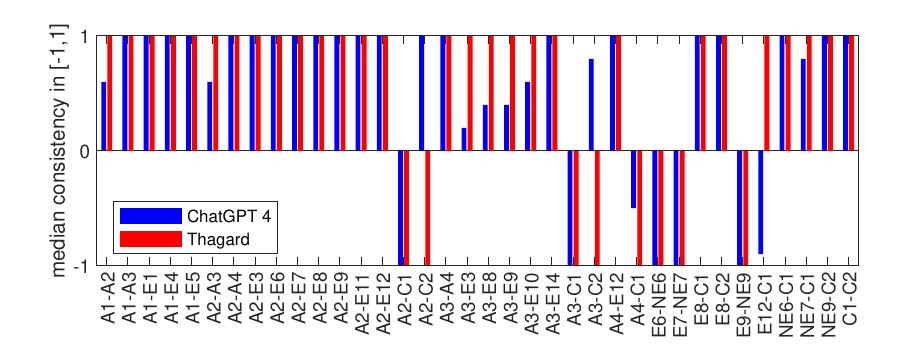}
  \caption{Median numerical consistency ratings from ChatGPT 4 versus the handcrafted consistency ratings in \cite{thagard1992adversarial}. 
  }
  \label{fig:vincennes}
\end{figure}

Another example (also reported in an appendix of \cite{huntsman2025neurosymbolic}) 
gives a further basis of confidence that even deprecated LLMs can outperform humans at producing consistency ratings. The paper \cite{thagard1992adversarial} includes an example of CDI modeling the decision problem facing the captain of the USS \emph{Vincennes} on 3 July 1988: was an aircraft taking off from the dual civilian-military airfield at Bandar Abbas a hostile F-14 about to attack the ship, or a civilian airliner? As in \cite{thagard1992adversarial}, we used the formal report on the incident \cite{fogarty1988formal}: specifically, we provided relevant parts of the preliminary statement and executive summary as background context to a prompt. Prompts also variously contained positive evidence (propositions E*, taken almost verbatim from \S III.C.1.b of \cite{fogarty1988formal}), negative evidence (propositions NE*), and a handful of hypotheses concerning the downed aircraft (``Track 4131''; attacking aircraft propositions A*, and commercial aircraft propositions C*). For an ``apples-to-apples'' analysis, we stipulated the graph structure in \cite{thagard1992adversarial} and rated the consistency of precisely the same pairs of propositions. The results are in Figure \ref{fig:vincennes}.

Here, we list the four largest divergences between the ratings in \cite{thagard1992adversarial} and the median ratings produced by ChatGPT 4. In each case, ChatGPT’s rating improves on the one in \cite{thagard1992adversarial}:
\begin{enumerate}
    \item A2-C2. These are obviously consistent (and recognized as such in \cite{thagard1992adversarial}, but treated as inconsistent for technical reasons):
    \begin{itemize}
        \item A2 = ``Track 4131 was an F-14.''
        \item C2 = ``Track 4131 was taking off.'' 
    \end{itemize}
    \item A3-E3. ChatGPT cited technical failures and misunderstandings as plausible:
    \begin{itemize}
        \item A3 = ``Track 4131 intended to attack.''
        \item E3 = ``Track 4131 was not responding to verbal warnings over [air distress frequencies].''
    \end{itemize}
    \item A3-C2. Both A3 and C2 are detailed above. These are obviously consistent (and recognized as such in \cite{thagard1992adversarial}, but treated as inconsistent for technical reasons).
    \item E12-C1. ChatGPT cited navigation and communications emissions of airliners as relevant:
    \begin{itemize}
        \item E12 = ``No [electronic emissions were reported] from track 4131, however, F-14s can fly [without electronic emissions].''
        \item C1 = ``Track 4131 was a commercial airliner.''
    \end{itemize}
\end{enumerate}

In short, ChatGPT 4 outperformed an expert human gauging consistency of propositions. It is likely that LLMs can also outperform humans at gauging the \emph{relevance} of sets of propositions. In fact, the attention mechanism that enables LLMs excels at things very much like this \cite{vaswani2017attention}. Some preliminary results that indicate that this idea may work, though not entirely straightforwardly. However, there is clear evidence that some LLMs can compute pairwise relevance and consistency of an entire set of propositions with superhuman fidelity from a single prompt \cite{huntsman2025neurosymbolic}. The sequel is the first of several examples in which we use a LLM to compile a coherence graph and perform CDI on the result.

\section{\label{sec:examples}Examples}

The examples in \S \ref{sec:oval} and \S \ref{sec:brown} below both feature ``external'' facts that inform which parts to accept and reject. In the former case, proposition $o$ in Table \ref{tab:oval} has a privileged epistemic status; in the latter case, propositions $p5$ and $p8$ in Table \ref{tab:brown} were famously demonstrated through a ``doll test.'' The other examples below, in \S \ref{sec:melian} and \S \ref{sec:inherit}, lack such external facts, and the distinction between acceptance and rejection is less straightforward \emph{a priori}. If there are not sufficient recognized data priorities to ground an acceptance/rejection decision, a user will have to review the output of CDI to break the symmetry.

\subsection{\label{sec:oval}28 February 2025 Oval Office meeting}

On 28 February 2025, US President Donald Trump, US Vice President J. D. Vance, and Ukrainian President Volodymyr Zelenskyy held a contentious meeting that was broadcast live from the Oval Office \cite{ap2025oval}. In a preliminary semiautomated instantiation of our approach performed shortly thereafter, we 
\begin{itemize}
    \item asked GPT-4o to produce a labeled list of substantive propositions derived from the meeting transcript;
    \item post-processed the list by, e.g., consolidating into 19 total propositions, replacing pronouns with antecedents, etc.;
    \item gave o3-mini the prompt in Table \ref{tab:prompt_practical} five times; \footnote{Since then, we usually take 25-50 prompt responses.}
    \item manually aligned graph edges across responses, keeping only edges that occurred most of the time and averaging the weights; \footnote{Since then, we have taken medians in place of averages: the former gives the optimal $L^1$/graph edit distance loss; and we can characterize this loss and convergence by considering subsamples as in \S \ref{sec:l1_distances}.}
    \item extracted the graph component on six of the 19 total vertices (labeled $\{a, e, k, o, p, r\}$) whose data related to war termination (two other connected components related to US presidents and to gratitude from Ukraine towards the US);
    \item quantized and rescaled the weights to be in $\{-1,-1/2,0,1/2,1\}$;
    \item computed the coherence of every bipartition.
\end{itemize}
The propositions relating to war termination are shown in Table \ref{tab:oval}. {\color{red}Red labels and highlights indicate subjective or arguably subjective language.} {\color{blue}The blue proposition $o$ does not contain such language (and for that matter, is objectively true \cite{budjeryn2020budapest}), thus imbuing any part of a maximally coherent bipartition it inhabits with priority.} The corresponding connected component of the quantized coherence graph is shown in Figure \ref{fig:oval}.

\begin{table}[htbp]
\caption{Six propositions derived from the meeting transcript in \cite{ap2025oval}.}
\begin{center}
\begin{tabular}{|l|}
\hline
{\color{red}- $a$:}	Effective diplomacy requires engaging both Russia and Ukraine, \\ \quad and overt hostility toward Vladimir Putin {\color{red}would} hinder negotiations. \\
{\color{red}- $e$:}	A ceasefire between Ukraine and Russia {\color{red}should} be pursued first, \\ \quad as a ceasefire is easier to achieve than a full peace agreement. \\
{\color{red}- $k$:}	Diplomacy between the United States, Ukraine, and Russia {\color{red}is the} \\ \quad {\color{red}best} way to achieve peace. \\
{\color{blue}- $o$:} {\color{blue}Russia has occupied Ukrainian territory, including Crimea and} \\ \quad {\color{blue}eastern Ukraine, since 2014 and has consistently broken agreements} \\ \quad {\color{blue}with Ukraine.} \\
{\color{red}- $p$:}	Diplomacy between Ukraine and Russia has been attempted \\ \quad through ceasefires and prisoner exchanges but has repeatedly failed \\ \quad {\color{red}due to} Russian violations. \\
{\color{red}- $r$:}	A ceasefire between Ukraine and Russia without guarantees is \\ \quad meaningless {\color{red}because} Russia has a history of breaking agreements \\ \quad with Ukraine. \\
\hline
\end{tabular}
\label{tab:oval}
\end{center}
\end{table}

\begin{figure}[htbp]
  \centering
    \resizebox{.5\columnwidth}{!}{%
        \begin{tikzpicture}
          \tikzset{
            node/.style={circle, draw=black, fill=white, minimum size=.25cm},
          }

          \node[node] (a) at (1, 1.732) {{\color{red}$a$}};
          \node[node] (e) at (2, 0) {{\color{red}$e$}};
          \node[node] (k) at (1, -1.732) {{\color{red}$k$}};
          \node[node] (o) at (-1, -1.732) {{\color{blue}$o$}};
          \node[node] (p) at (-2, 0) {{\color{red}$p$}};
          \node[node] (r) at (-1, 1.732) {{\color{red}$r$}};
        
          \draw[solid, color={rgb,1:red,0.000;green,0.000;blue,1.000}, opacity=1.00, line width=3pt] (a) -- (e);
          \draw[solid, color={rgb,1:red,0.000;green,0.000;blue,1.000}, opacity=1.00, line width=3pt] (a) -- (k);
          \draw[solid, color={rgb,1:red,0.250;green,0.000;blue,0.750}, opacity=0.50, line width=3pt] (a) -- (p);
          \draw[solid, color={rgb,1:red,0.250;green,0.000;blue,0.750}, opacity=0.50, line width=3pt] (a) -- (r);

          \draw[solid, color={rgb,1:red,0.000;green,0.000;blue,1.000}, opacity=1.00, line width=3pt] (e) -- (k);
          \draw[dashed, color={rgb,1:red,1.000;green,0.000;blue,0.000}, opacity=1.00, line width=3pt] (e) -- (p);
          \draw[dashed, color={rgb,1:red,1.000;green,0.000;blue,0.000}, opacity=1.00, line width=3pt] (e) -- (r);

          \draw[solid, color={rgb,1:red,0.250;green,0.000;blue,0.750}, opacity=0.50, line width=3pt] (k) -- (r);
          \draw[dashed, color={rgb,1:red,1.000;green,0.000;blue,0.000}, opacity=1.00, line width=3pt] (k) -- (p);

          \draw[solid, color={rgb,1:red,0.000;green,0.000;blue,1.000}, opacity=1.00, line width=3pt] (o) -- (p);
          \draw[solid, color={rgb,1:red,0.000;green,0.000;blue,1.000}, opacity=1.00, line width=3pt] (o) -- (r);

          \draw[solid, color={rgb,1:red,0.000;green,0.000;blue,1.000}, opacity=1.00, line width=3pt] (p) -- (r);

        \end{tikzpicture}
    }
  \caption{A quantized coherence graph formed from the propositions in Table \ref{tab:oval}. Intermediate weights (here all equal to $+0.5$) are indicated with intermediate color {\color{rgb,1:red,0.250;green,0.000;blue,0.750}(here, all bluish purple)} and partial transparency along with the line style (solid or dashed) corresponding to the sign of the weights.
  }
  \label{fig:oval}
\end{figure}

The most coherent of the $2^{6-1}$ bipartitions is $\{\{a,e,k\},\{o,p,r\}\}$. That is, the maximum cut is straight down the middle of the graph drawing, with a coherence of $1.5$, since it cuts {\color{red}three edges with weight -1 (i.e., $(e,p)$, $(e,r)$, and $(k,p)$)} and {\color{rgb,1:red,0.250;green,0.000;blue,0.750}three edges with weight +0.5 (i.e., $(a,p)$, $(a,r)$, and $(k,r)$)}. Table \ref{tab:bipartitions} shows the coherence of some more bipartitions. Meanwhile, of the six underlying propositions, only proposition $o$ is not speculative (note that this is not automatically the same as not false). If we accept proposition $o$ as an observation, then this breaks the symmetry between acceptance and rejection on bipartitions. 

\begin{table}[htbp]
\caption{Some bipartitions of the graph in Figure \ref{fig:oval} and their coherence.}
\begin{center}
\begin{tabular}{|c|c|}
\hline
\textbf{Bipartition} & \textbf{Coherence} \\
\hline
$\{\{o,p,r\},\{a,e,k\}\}$ & 1.5 (best) \\
\hline
$\{\{a,o,p,r\},\{e,k\}\}$ & 0.5 \\
\hline
$\{\{a,e,k,o,p,r\},\varnothing\}$ & 0.0 \\
\hline
$\{\{a,k,o,p,r\},\{e\}\}$ & 0.0 \\
\hline
$\{\{a,e,k,o\},\{p,r\}\}$ & -0.5 \\
\hline
$\{\{a,e,k,o,r\},\{p\}\}$ & -0.5 \\
\hline
$\{\{o,p\},\{a,e,k,r\}\}$ & -0.5 \\
\hline
$\{\{e,o,p,r\},\{a,k\}\}$ & -2.5 \\
\hline
$\{\{a,o\},\{e,k,p,r\}\}$ & -5 (worst) \\
\hline
\end{tabular}
\label{tab:bipartitions}
\end{center}
\end{table}

That is, a ``classical'' principle of data priority in CDI weighs in favor of accepting $\{o,p,r\}$ and rejecting $\{a,e,k\}$ instead of the other way around. In particular, CDI leads to acceptance of proposition $r$:
\begin{quote}
    A ceasefire between Ukraine and Russia without guarantees is meaningless because Russia has a history of breaking agreements with Ukraine.
\end{quote}

It is and was obvious that this example could be completely automated. Some examples of a fully automated approach are outlined in the following sections.

\subsection{\label{sec:melian}The Melian Dialogue}

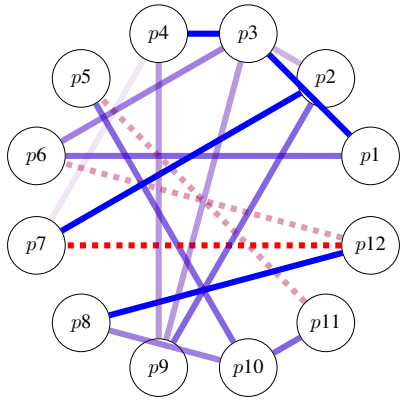
\begin{figure}[htbp]
  \centering
    \resizebox{.6\columnwidth}{!}{%
        \begin{tikzpicture}
          \tikzset{
            node/.style={circle, draw=black, fill=white, minimum size=1cm},
          }
        
          \node[node] (nodep1) at (2.8978, .77645) {$p$1};
          \node[node] (nodep10) at (.77645, -2.8978) {$p$10};
          \node[node] (nodep11) at (2.1213, -2.1213) {$p$11};
          \node[node] (nodep12) at (2.8978, -.77645) {$p$12};
          \node[node] (nodep2) at (2.1213, 2.1213) {$p$2};
          \node[node] (nodep3) at (.77645, 2.8978) {$p$3};
          \node[node] (nodep4) at (-.77645, 2.8978) {$p$4};
          \node[node] (nodep5) at (-2.1213, 2.1213) {$p$5};
          \node[node] (nodep6) at (-2.8978, .77645) {$p$6};
          \node[node] (nodep7) at (-2.8978, -.77645) {$p$7};
          \node[node] (nodep8) at (-2.1213, -2.1213) {$p$8};
          \node[node] (nodep9) at (-.77645, -2.8978) {$p$9};
        
          \draw[solid, color={rgb,1:red,0.500;green,0.000;blue,0.500}, opacity=0.00, line width=3pt] (nodep1) -- (nodep10);
          \draw[solid, color={rgb,1:red,0.500;green,0.000;blue,0.500}, opacity=0.00, line width=3pt] (nodep1) -- (nodep11);
          \draw[solid, color={rgb,1:red,0.500;green,0.000;blue,0.500}, opacity=0.00, line width=3pt] (nodep1) -- (nodep12);
          \draw[solid, color={rgb,1:red,0.500;green,0.000;blue,0.500}, opacity=0.00, line width=3pt] (nodep1) -- (nodep2);
          \draw[solid, color={rgb,1:red,0.000;green,0.000;blue,1.000}, opacity=1.00, line width=3pt] (nodep1) -- (nodep3);
          \draw[solid, color={rgb,1:red,0.500;green,0.000;blue,0.500}, opacity=0.00, line width=3pt] (nodep1) -- (nodep4);
          \draw[solid, color={rgb,1:red,0.500;green,0.000;blue,0.500}, opacity=0.00, line width=3pt] (nodep1) -- (nodep5);
          \draw[solid, color={rgb,1:red,0.200;green,0.000;blue,0.800}, opacity=0.60, line width=3pt] (nodep1) -- (nodep6);
          \draw[solid, color={rgb,1:red,0.500;green,0.000;blue,0.500}, opacity=0.00, line width=3pt] (nodep1) -- (nodep7);
          \draw[solid, color={rgb,1:red,0.500;green,0.000;blue,0.500}, opacity=0.00, line width=3pt] (nodep1) -- (nodep8);
          \draw[solid, color={rgb,1:red,0.500;green,0.000;blue,0.500}, opacity=0.00, line width=3pt] (nodep1) -- (nodep9);
          \draw[solid, color={rgb,1:red,0.200;green,0.000;blue,0.800}, opacity=0.60, line width=3pt] (nodep10) -- (nodep11);
          \draw[solid, color={rgb,1:red,0.500;green,0.000;blue,0.500}, opacity=0.00, line width=3pt] (nodep10) -- (nodep12);
          \draw[solid, color={rgb,1:red,0.500;green,0.000;blue,0.500}, opacity=0.00, line width=3pt] (nodep10) -- (nodep2);
          \draw[solid, color={rgb,1:red,0.500;green,0.000;blue,0.500}, opacity=0.00, line width=3pt] (nodep10) -- (nodep3);
          \draw[solid, color={rgb,1:red,0.500;green,0.000;blue,0.500}, opacity=0.00, line width=3pt] (nodep10) -- (nodep4);
          \draw[solid, color={rgb,1:red,0.200;green,0.000;blue,0.800}, opacity=0.60, line width=3pt] (nodep10) -- (nodep5);
          \draw[solid, color={rgb,1:red,0.500;green,0.000;blue,0.500}, opacity=0.00, line width=3pt] (nodep10) -- (nodep6);
          \draw[solid, color={rgb,1:red,0.500;green,0.000;blue,0.500}, opacity=0.00, line width=3pt] (nodep10) -- (nodep7);
          \draw[solid, color={rgb,1:red,0.250;green,0.000;blue,0.750}, opacity=0.50, line width=3pt] (nodep10) -- (nodep8);
          \draw[solid, color={rgb,1:red,0.500;green,0.000;blue,0.500}, opacity=0.00, line width=3pt] (nodep10) -- (nodep9);
          \draw[solid, color={rgb,1:red,0.500;green,0.000;blue,0.500}, opacity=0.00, line width=3pt] (nodep11) -- (nodep12);
          \draw[solid, color={rgb,1:red,0.500;green,0.000;blue,0.500}, opacity=0.00, line width=3pt] (nodep11) -- (nodep2);
          \draw[solid, color={rgb,1:red,0.500;green,0.000;blue,0.500}, opacity=0.00, line width=3pt] (nodep11) -- (nodep3);
          \draw[solid, color={rgb,1:red,0.500;green,0.000;blue,0.500}, opacity=0.00, line width=3pt] (nodep11) -- (nodep4);
          \draw[dashed, color={rgb,1:red,0.700;green,0.000;blue,0.300}, opacity=0.40, line width=3pt] (nodep11) -- (nodep5);
          \draw[solid, color={rgb,1:red,0.500;green,0.000;blue,0.500}, opacity=0.00, line width=3pt] (nodep11) -- (nodep6);
          \draw[solid, color={rgb,1:red,0.500;green,0.000;blue,0.500}, opacity=0.00, line width=3pt] (nodep11) -- (nodep7);
          \draw[solid, color={rgb,1:red,0.500;green,0.000;blue,0.500}, opacity=0.00, line width=3pt] (nodep11) -- (nodep8);
          \draw[solid, color={rgb,1:red,0.500;green,0.000;blue,0.500}, opacity=0.00, line width=3pt] (nodep11) -- (nodep9);
          \draw[solid, color={rgb,1:red,0.500;green,0.000;blue,0.500}, opacity=0.00, line width=3pt] (nodep12) -- (nodep2);
          \draw[solid, color={rgb,1:red,0.500;green,0.000;blue,0.500}, opacity=0.00, line width=3pt] (nodep12) -- (nodep3);
          \draw[solid, color={rgb,1:red,0.500;green,0.000;blue,0.500}, opacity=0.00, line width=3pt] (nodep12) -- (nodep4);
          \draw[solid, color={rgb,1:red,0.500;green,0.000;blue,0.500}, opacity=0.00, line width=3pt] (nodep12) -- (nodep5);
          \draw[dashed, color={rgb,1:red,0.700;green,0.000;blue,0.300}, opacity=0.40, line width=3pt] (nodep12) -- (nodep6);
          \draw[dashed, color={rgb,1:red,1.000;green,0.000;blue,0.000}, opacity=1.00, line width=3pt] (nodep12) -- (nodep7);
          \draw[solid, color={rgb,1:red,0.000;green,0.000;blue,1.000}, opacity=1.00, line width=3pt] (nodep12) -- (nodep8);
          \draw[solid, color={rgb,1:red,0.500;green,0.000;blue,0.500}, opacity=0.00, line width=3pt] (nodep12) -- (nodep9);
          \draw[solid, color={rgb,1:red,0.350;green,0.000;blue,0.650}, opacity=0.30, line width=3pt] (nodep2) -- (nodep3);
          \draw[solid, color={rgb,1:red,0.500;green,0.000;blue,0.500}, opacity=0.00, line width=3pt] (nodep2) -- (nodep4);
          \draw[solid, color={rgb,1:red,0.500;green,0.000;blue,0.500}, opacity=0.00, line width=3pt] (nodep2) -- (nodep5);
          \draw[solid, color={rgb,1:red,0.500;green,0.000;blue,0.500}, opacity=0.00, line width=3pt] (nodep2) -- (nodep6);
          \draw[solid, color={rgb,1:red,0.000;green,0.000;blue,1.000}, opacity=1.00, line width=3pt] (nodep2) -- (nodep7);
          \draw[solid, color={rgb,1:red,0.500;green,0.000;blue,0.500}, opacity=0.00, line width=3pt] (nodep2) -- (nodep8);
          \draw[solid, color={rgb,1:red,0.200;green,0.000;blue,0.800}, opacity=0.60, line width=3pt] (nodep2) -- (nodep9);
          \draw[solid, color={rgb,1:red,0.000;green,0.000;blue,1.000}, opacity=1.00, line width=3pt] (nodep3) -- (nodep4);
          \draw[solid, color={rgb,1:red,0.500;green,0.000;blue,0.500}, opacity=0.00, line width=3pt] (nodep3) -- (nodep5);
          \draw[solid, color={rgb,1:red,0.250;green,0.000;blue,0.750}, opacity=0.50, line width=3pt] (nodep3) -- (nodep6);
          \draw[solid, color={rgb,1:red,0.500;green,0.000;blue,0.500}, opacity=0.00, line width=3pt] (nodep3) -- (nodep7);
          \draw[solid, color={rgb,1:red,0.500;green,0.000;blue,0.500}, opacity=0.00, line width=3pt] (nodep3) -- (nodep8);
          \draw[solid, color={rgb,1:red,0.300;green,0.000;blue,0.700}, opacity=0.40, line width=3pt] (nodep3) -- (nodep9);
          \draw[solid, color={rgb,1:red,0.500;green,0.000;blue,0.500}, opacity=0.00, line width=3pt] (nodep4) -- (nodep5);
          \draw[solid, color={rgb,1:red,0.500;green,0.000;blue,0.500}, opacity=0.00, line width=3pt] (nodep4) -- (nodep6);
          \draw[solid, color={rgb,1:red,0.450;green,0.000;blue,0.550}, opacity=0.10, line width=3pt] (nodep4) -- (nodep7);
          \draw[solid, color={rgb,1:red,0.500;green,0.000;blue,0.500}, opacity=0.00, line width=3pt] (nodep4) -- (nodep8);
          \draw[solid, color={rgb,1:red,0.300;green,0.000;blue,0.700}, opacity=0.40, line width=3pt] (nodep4) -- (nodep9);
          \draw[solid, color={rgb,1:red,0.500;green,0.000;blue,0.500}, opacity=0.00, line width=3pt] (nodep5) -- (nodep6);
          \draw[solid, color={rgb,1:red,0.500;green,0.000;blue,0.500}, opacity=0.00, line width=3pt] (nodep5) -- (nodep7);
          \draw[solid, color={rgb,1:red,0.500;green,0.000;blue,0.500}, opacity=0.00, line width=3pt] (nodep5) -- (nodep8);
          \draw[solid, color={rgb,1:red,0.500;green,0.000;blue,0.500}, opacity=0.00, line width=3pt] (nodep5) -- (nodep9);
          \draw[solid, color={rgb,1:red,0.500;green,0.000;blue,0.500}, opacity=0.00, line width=3pt] (nodep6) -- (nodep7);
          \draw[solid, color={rgb,1:red,0.500;green,0.000;blue,0.500}, opacity=0.00, line width=3pt] (nodep6) -- (nodep8);
          \draw[solid, color={rgb,1:red,0.500;green,0.000;blue,0.500}, opacity=0.00, line width=3pt] (nodep6) -- (nodep9);
          \draw[solid, color={rgb,1:red,0.500;green,0.000;blue,0.500}, opacity=0.00, line width=3pt] (nodep7) -- (nodep8);
          \draw[solid, color={rgb,1:red,0.500;green,0.000;blue,0.500}, opacity=0.00, line width=3pt] (nodep7) -- (nodep9);
          \draw[solid, color={rgb,1:red,0.500;green,0.000;blue,0.500}, opacity=0.00, line width=3pt] (nodep8) -- (nodep9);
        \end{tikzpicture}
    }
  \caption{The median of 30 coherence graphs produced by single prompts to o1-mini along the lines of \S \ref{sec:prompt_practical}. Convergence of the median coherence graph is illustrated in \S \ref{sec:l1_distances}. {\color{blue}Consistent (solid blue)} and {\color{red}inconsistent (red dashed)} pairs of vertices (= propositions) respectively get {\color{blue}positive} and {\color{red}negative} weights, with intermediate colors and transparency indicating the strength of (in)consistency. The optimal cut has parts $\{p1, p2, p3, p4, p6, p7, p9\}$ and $\{p5, p8, p10, p11, p12\}$. 
  }
  \label{fig:melianDialogue}
\end{figure}

The Melian Dialogue in \cite{thucydides1874history} is a famous argument in which the Athenians tell the Melians that ``the strong [Athenians] do what they can and the weak [Melians] suffer what they must.'' Table \ref{tab:melianDialogue} gives the 12 most important propositions from the Dialogue as extracted and processed by GPT-4o. (NB. Lacedaemon is an ancient name for Sparta.) The table also shows the maximum cut in a LLM-compiled coherence graph formed from these propositions that is shown in Figure \ref{fig:melianDialogue}.

The graph and its maximum cut indicates that the Athenian proposition $p$5 [hope is risky and unreliable, especially for those with few resources like the Melians] actually coheres with the Melian position; likewise, the Melian proposition $p$9 [Athenian aggression risks creating more enemies for Athens and alienating neutral states] actually coheres with the Athenian position. History shows the significance of these propositions. A hopeful Melos was brutally conquered by Athens in 416 BCE, during a truce with Sparta. Sparta later defeated Athens in the Peloponnesian War in 405 BCE, removed Athenians from Melos, and resettled surviving Melians there. 

\begin{figure}[htbp]
  \centering
    \resizebox{\columnwidth}{!}{%
      \includegraphics[trim = 0mm 0mm 0mm 0mm, clip, width=\columnwidth,keepaspectratio]{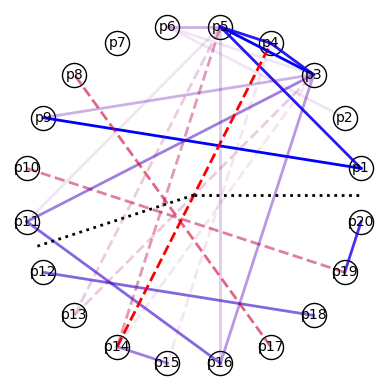}
    }
    \caption{The median coherence graph constructed by o1-mini over the propositions in Table \ref{tab:melian}. A black dashed line indicates the separation between participants in the argument, not (necessarily) an optimal cut. Here, an optimal cut groups $p$16 (and optionally, both $p$12 and $p$18) with the Athenian propositions $p$1-$p$11 (optionally, without $p$7).
  }
  \label{fig:melian}
\end{figure}

Table \ref{tab:melian} and Figure \ref{fig:melian} give a larger example suggesting that the Melians would have done well to focus on the proposition $p$16 [trying all options before submitting is important] to delay conflict.

\begin{table}[htbp]
\caption{A small set of propositions summarizing the Melian Dialogue.}
\begin{center}
\begin{tabular}{|l|}
\hline
{\bf \# Athenians} \\
- $p$1: The strong dominate and the weak endure. \\
- $p$2: Internal rebellion is a greater threat to Athens than external \\ \quad rivals like Lacedaemon. \\
- $p$3: Subjugating Melos would strengthen Athenian security and \\ \quad expand the Athenian empire. \\
- $p$4: Melian neutrality would show Athenian weakness. \\
- $p$6: Both gods and men support ruling wherever possible. \\
- $p$7: The Lacedaemonians prioritize security and are unlikely to \\ \quad intervene in support of Melos. \\

\ \\

{\bf \# Melians} \\
- $p$9: Athenian aggression risks creating more enemies for Athens \\ \quad and alienating neutral states. \\

\ \\

{\bf *** MAXIMUM CUT IS HERE ***} \\

\ \\

{\bf \# Athenians} \\
- $p$5: Hope is risky and unreliable, especially for those with few \\ \quad resources like the Melians. \\

\ \\

{\bf \# Melians} \\
- $p$8: Justice and fairness protect weaker states like Melos. \\
- $p$10: Resisting Athens is about honor and survival for the Melians. \\
- $p$11: War's unpredictability offers hope to the Melians against \\ \quad stronger foes like Athens. \\
- $p$12: The gods and the Lacedaemonians will support Melos due to \\ \quad justice and kinship with the Melians. \\
\hline
\end{tabular}
\label{tab:melianDialogue}
\end{center}
\end{table}

\begin{table}[htbp]
\caption{A larger set of propositions summarizing the Melian Dialogue.}
\begin{center}
\begin{tabular}{|l|}
\hline
{\bf \# Athenians} \\
- $p$1: Right is determined by power; the strong dominate, and the \\ \quad weak endure. \\
- $p$2: Internal rebellion is more concerning than external threats. \\
- $p$3: Preserving the Athenian empire and Melos's submission is \\ \quad beneficial. \\
- $p$4: Neutrality from Melos is unacceptable as it would show \\ \quad Athenian weakness. \\
- $p$5: Subjugating Melos would strengthen Athenian security and \\ \quad the Athenian empire. \\
- $p$6: Islanders like Melos are more threatening than continentals. \\
- $p$7: The conflict with Melos is about survival, not honor. \\
- $p$8: Hope is unreliable and can lead to ruin. \\
- $p$9: Athenian actions align with the natural order of ruling where \\ \quad possible. \\
- $p$10: Lacedaemonian support for Melos is doubtful due to the \\ \quad Lacedaemonians' cautious nature. \\
- $p$11: Submitting to Athens is a secure and honorable choice for \\ \quad Melos. \\

\ \\

{\bf \# Melians} \\

- $p$16: Trying all options before submitting is important. \\

\ \\

{\bf *** A MAXIMUM CUT IS HERE ***} \\

\ \\

{\bf \# Melians} \\
- $p$12: Fairness and justice are crucial in conflict. \\
- $p$18: Divine support exists for the just cause. \\

\ \\

{\bf *** ANOTHER MAXIMUM CUT IS HERE ***} \\

\ \\

{\bf \# Melians} \\
- $p$13: Destroying Melos would set a dangerous precedent for \\ \quad Athens. \\
- $p$14: Neutrality is preferable to choosing sides. \\
- $p$15: Attacking neutrals could create more enemies for Athens. \\
- $p$17: War's unpredictability offers hope. \\
- $p$19: Lacedaemonian aid is expected due to kinship and honor. \\
- $p$20: Proximity to Peloponnese makes Lacedaemonian help more \\ \quad feasible. \\
\hline
\end{tabular}
\label{tab:melian}
\end{center}
\end{table}

\subsection{\label{sec:inherit}\emph{Inherit the Wind}}

\begin{figure}[htbp]
  \centering
    \resizebox{\columnwidth}{!}{%
      \includegraphics[trim = 0mm 0mm 0mm 0mm, clip, width=\columnwidth,keepaspectratio]{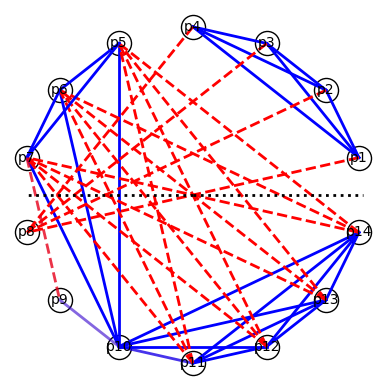}
    }
    \caption{The median coherence graph constructed by o1-mini from the propositions in Table \ref{tab:inheritProps}. 
  }
  \label{fig:inherit}
\end{figure}

The play \emph{Inherit the Wind} \cite{lawrence2000inherit} is---along with the movie of the same name---a fictionalization of the famous  ``Scopes monkey trial'' involving a Tennessee state law that prohibited teaching the theory of evolution in public schools. Table \ref{tab:inheritProps} lists the most important propositions in the climactic Act II, Scene 2 of \cite{lawrence2000inherit} as extracted by GPT-4o, along with the maximum cut in the corresponding coherence graph shown in Figure \ref{fig:inherit}. 

This account using CDI correctly separates arguments by both counsels. 
\footnote{
It is worth noting that CDI shows that the beliefs in the key debate are drawn in much sharper contrast than in the Melian Dialogue.
}
CDI also correctly indicates that proposition $p$10 is the pivot of the entire play. Following the exchange in Table \ref{tab:inheritScene}, Drummond proceeds to cast doubt on proposition $p$10 \emph{en route} to completely undermining the rest of Brady's arguments.

\begin{table}[htbp]
\caption{Propositions derived from Act II, Scene 2 of \cite{lawrence2000inherit}.}
\begin{center}
\begin{tabular}{|l|}
\hline
{\bf \# Drummond} \\ 
- $p$1: Expert testimony on evolution should be included in the trial of \\ \quad Bertram Cates. \\
- $p$2: Understanding evolution is essential for the jury's judgment in \\ \quad the trial. \\
- $p$3: The exclusion of scientific experts from the trial is wrong. \\
- $p$4: Bertram Cates' teachings on evolution are not criminal. \\
- $p$5: Matthew Harrison Brady's literal interpretation of the Bible is \\ \quad flawed. \\
- $p$6: Reason and scientific progress are more important than strict \\ \quad adherence to religious texts. \\
- $p$7: The Bible is not the sole source of truth and should not be the \\ \quad only reference in matters of education and law. \\

\ \\

{\bf *** MAXIMUM CUT IS HERE ***} \\

\ \\

{\bf \# Brady} \\
- $p$8: Testimony on evolution should not be included in the trial of \\ \quad Bertram Cates. \\
- $p$9: The law excludes evolution from being taught in schools. \\
- $p$10: Brady is an authority on the Bible but has not read Charles \\ \quad Darwin's ``Origin of Species.'' \\
- $p$11: The Bible's literal truth is valid, and God can change natural \\ \quad law as described in the Bible. \\ 
- $p$12: Scientific evidence that contradicts the Bible's teachings is \\ \quad invalid. \\
- $p$13: Faith in the Bible is more important than scientific reasoning \\ \quad and evidence. \\
- $p$14: God guides Brady's actions, and opposing Brady is akin to \\ \quad opposing God. \\
\hline
\end{tabular}
\label{tab:inheritProps}
\end{center}
\end{table}

\begin{table}[htbp]
\caption{The exchange of \cite{lawrence2000inherit} in the key court scene (Act II, Scene 2).}
\begin{center}
\begin{tabular}{|l|}
\hline
DRUMMOND \\ 
\emph{(Scowling)} \\ 
\quad In other words, the court rules out any expert testimony on Charles \\ \quad Darwin’s \emph{Origin of Species} or \emph{Descent of Man?} \\ 
JUDGE \\ 
\quad The court so rules. \\ 
\emph{(DRUMMOND is flabbergasted. His case is cooked and he knows it.} \\
\emph{He looks around helplessly.)} \\ 
DRUMMOND \\ 
\emph{(There’s the glint of an idea in his eye.)} \\ 
\quad Would the court admit expert testimony regarding a book known \\ \quad as the Holy Bible? \\ 
JUDGE \\ 
\emph{(Hesitates, turns to BRADY)} \\ 
\quad Any objection, Colonel Brady? \\ 
BRADY \\ 
\quad If the counsel can advance the case of the defendant through the \\ \quad use of the Holy Scriptures, the prosecution will take no exception! \\ 
DRUMMOND \\ 
\quad Good! (With relish) I call to the stand one of the world’s foremost \\ \quad experts on the Bible and its teachings – Matthew Harrison Brady! \\ 
\emph{(There is an uproar in the courtroom. The JUDGE raps for order.)} \\ 
\hline
\end{tabular}
\label{tab:inheritScene}
\end{center}
\end{table}

\subsection{\label{sec:brown}\emph{Brown v. Board of Education}}

\begin{figure}[htbp]
  \centering
    \resizebox{\columnwidth}{!}{%
      \includegraphics[trim = 0mm 0mm 0mm 0mm, clip, width=\columnwidth,keepaspectratio]{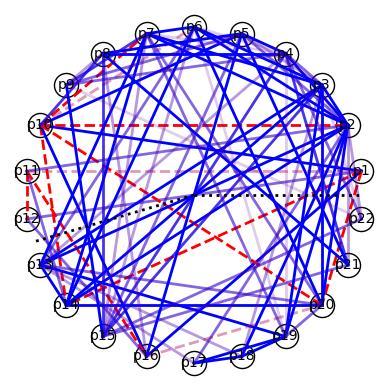}
    }
    \caption{The median coherence graph constructed by o1-mini from the propositions in Table \ref{tab:brown}.
  }
  \label{fig:brown}
\end{figure}

The faultline in the landmark Supreme Court decision in \emph{Brown v. Board of Education} was between the segregationist doctrine of ``separate but equal'' from \emph{Plessy v. Ferguson} and social harm demonstrated by a ``doll test'' \cite{oyez2025brown}. As Table \ref{tab:brown} and Figure \ref{fig:brown} illustrate, CDI exposes this faultline. The maximum cut shown in Table \ref{tab:brown} indicates a resolution that incorporates the losing counsel Paul Wilson's own arguments $p$15, $p$16, and $p$21. Indeed, as the commentary in \cite{oyez2025brown} points out:
\begin{quote}
Wilson's argument quickly boiled down to three things: first, school boards in Kansas had the right to make their own decisions regarding separate facilities for Black students; second, as long as schools were equal in every measurable respect, the school board was not violating the Constitution [corresponding to $p$15 in Figure \ref{fig:brown}]; and third, there just weren't enough Black students in the state to even make it a big deal [corresponding to $p$21 in Figure \ref{fig:brown}] \dots Wilson focused on a question that would become central to the \emph{Brown} decision: Does segregation itself cause harm to Black children? 
\end{quote}

\begin{table}[htbp]
\caption{Propositions derived from oral argument in \emph{Brown v. Board}.}
\begin{center}
\begin{tabular}{|l|}
\hline
{\bf \# Robert L. Carter} \\
- $p$1: The Kansas statute allowing segregated schools violates the \\ \quad Fourteenth Amendment. \\
- $p$4: The appellants must attend segregated elementary schools \\ \quad because of their race. \\
- $p$5: Segregation denies equal educational opportunities and harms \\ \quad the development of Negro children. \\
- $p$8: Segregation makes educational opportunities for Negro children \\ \quad inferior to those for white children. \\
- $p$9: There is no difference in physical facilities between schools for \\ \quad Negro children and schools for white children. \\
- $p$10: Segregation itself is unconstitutional under the Fourteenth \\ \quad Amendment. \\
- $p$12: \emph{Plessy v. Ferguson} does not apply to education. \\

\ \\

{\bf \# Paul Wilson} \\
- $p$15: There is no substantial inequality in educational facilities \\ \quad between schools for Negro children and schools for white children. \\
- $p$16: The "separate but equal" doctrine may not be valid. \\
- $p$21: Segregation is detrimental to Negro children but legally \\ \quad insignificant. \\

\ \\

{\bf *** MAXIMUM CUT IS HERE ***} \\

\ \\

{\bf \# Robert L. Carter} \\
- $p$2: The Kansas statute permits segregation in elementary schools, \\ \quad and in high schools only in Kansas City. \\
- $p$3: Kansas law prohibits racial distinctions in public schools \\ \quad without the Kansas statute. \\
- $p$6: Segregation complies with Kansas state law. \\
- $p$7: Kansas can impose racial distinctions if educational facilities \\ \quad are equal. \\
- $p$11: \emph{Plessy v. Ferguson} and \emph{Gong Lum v. Rice} require upholding \\ \quad the Kansas statute. \\

\ \\

{\bf \# Paul Wilson} \\
- $p$13: The Kansas statute allows, but does not require, segregation \\ \quad in cities with populations over 15,000. \\
- $p$14: The statute is constitutional, supported by state and federal \\ \quad court decisions. \\
- $p$17: Kansas has a small Negro population, less than four percent, \\ \quad mostly in urban areas. \\
- $p$18: Segregated schools exist in only nine cities in Kansas. \\
- $p$19: The Kansas statute was designed to let communities adjust to \\ \quad local conditions. \\
- $p$20: The Kansas Supreme Court has upheld the statute's \\ \quad constitutionality. \\
- $p$22: The appellants have not shown specific harm from attending \\ \quad segregated schools. \\
\hline
\end{tabular}
\label{tab:brown}
\end{center}
\end{table}

Wilson said in oral argument that 
\begin{quote}
this Court can overrule the \emph{Gong Lum} doctrine and the \emph{Plessy} doctrine, but nevertheless, until those cases are overruled, they are the best guide we have.    
\end{quote}
This remark corresponds directly to the proposition $p$16 in Figure \ref{fig:brown}. In short, CDI mirrors the reasoning of the court.

\section{\label{sec:law}(In)coherence in law}

Having demonstrated automatic CDI on several arguments, we shift focus for the remainder of the paper to the context of (in)coherence in law.

\subsection{\label{sec:incoherent}Law is incoherent}

Over 50 years ago, the development of sentencing guidelines was spurred by the analysis of \cite{frankel1973criminal}, which pointed out that:

\begin{quote}
[T]he evidence is conclusive that judges of widely varying attitudes on sentencing, administering statutes that confer huge measures of discretion, mete out widely divergent sentences where the divergences are explainable only by the variations among the judges, not by material differences in the defendants or their crimes.
\end{quote}

Recent evidence indicates that these sentencing guidelines are still applied incoherently \cite{topaz2025presidential}. Moreover, since the Supreme Court overturned \emph{Chevron} deference (i.e., the doctrine that courts deferred to federal agencies when interpreting statutes) in \emph{Loper Bright v. Raimondo}, it also seems likely that jurisprudential incoherence in administrative law will metastasize \cite{sunstein2024consequences}. Though this sea change may reduce temporal inconsistency \cite{dotan2005making}, in our view this is a dubious goal over sufficiently long timescales, since policies and interpretations must evolve over time to be responsive to circumstances. In any event, the goal of reducing temporal inconsistency may be thwarted anyway by tension between the surviving applicable doctrines. 

At least one mathematical model shows how incoherence can naturally arise via multi-step reasoning using abstract subjective criteria that depend on objective facts \cite{duck2022explaining}. At the same time, judges seem to behave more incoherently than juries, and there is an obvious explanation: a sample size of one admits more variation than a sample size of twelve.

This state of affairs has motivated inquiries into the nature of coherence in law and also---perhaps out of a sense of defeatism---substantive defenses of incoherence in law. Arguments from pluralism, ambiguity, and discretion have variously been advanced in this direction, with prominent themes that equality before the law need not amount to equity among outcomes or that laws must be applied consistently over time \cite{coons1987consistency,raz1992relevance,foran2022cornerstone,kress2010coherence}.

\subsection{\label{sec:worthy}Coherence is a worthy goal}

In our view, all of the arguments defending legal incoherence are hollow nods to expediency at the expense of fairness. Arguments against \emph{res judicata} or estoppel\footnote{See \url{https://www.law.cornell.edu/wex/res_judicata} and \url{https://www.law.cornell.edu/wex/estoppel}.} seem less likely than those against coherence precisely because the practical grounding of the former doctrines makes them expedient. Legal coherence threatens the fabric of any incoherent social construct: this is potentially good, but certainly dangerous. While long-term and moral considerations argue for fairness, short-term and social considerations argue for expediency. In a sense the ubiquitous Hart-Dworkin debate is about whether or not law should be grounded only socially or also morally \cite{shapiro2007hart}. 

Adherents of coherence can be expected to take a position closer to Dworkin. Others in this camp include Mark Elliott, who has argued for consistency as a ``free-standing principle'' of law \cite{elliott2018consistency}, and Alexander Peczenik, who said that the basic ideas of a coherence theory of law are ``reasonable support and weighing of reasons. All the rest is commentary.'' In particular, ``the law is what the most coherent theory of everything says it is'' \cite{peczenik2009law}.

Perhaps most comprehensively, Amalia Amaya has argued in favor of a coherentist approach to law, but only when grounded in epistemic virtues like diligence, courage to face criticism, perseverance in reasoning, and open-mindedness \cite{amaya2015tapestry,amaya2018coherence}. The basic idea is that an epistemically virtuous person will naturally recalibrate their beliefs to cohere with their observations in a process akin to John Rawls’ notion of reflective equilibrium \cite{rawls1951outline} or similarly, the systematic purging of any cognitive dissonance. As \cite{simon2004third} puts it:

\begin{quote}
    Coherence-based reasoning posits that the mind shuns cognitively complex and difficult decision tasks by reconstructing them into easy ones, yielding strong, confident conclusions. 
\end{quote}

In hard cases, this dynamical convergence to coherence may never arrive at a goal that changes in the face of new information and circumstances. However, a coherent approximation of truth can be broadened by incorporating additional coherent observations, and deepened by incorporating additional coherent beliefs \cite{thagard2007coherence}. In fact, this process helps explain the dynamics of scientific revolutions \cite{thagard2018conceptual}.

The coherentist view is comprehensively informed by cognitive science and experimentally supported by psychological case studies involving legal assessments \cite{holyoak1999bidirectional,simon2004third}. It is preferable to alternative models of legal inference on both psychological and computational grounds \cite{thagard2004causal} and well-suited for making decisions about ill-structured problems \cite{frigotto2015explanatory}, including as a prospective decision support tool for judges \cite{savelka2013coherence}. Coherence methods have been used to reach deliberative consensus \cite{joseph2009coherence} and solve normative inconsistencies \cite{criado2016coherence}. They can explicitly incorporate ethical considerations while reasoning explainably \cite{yilmaz2017computational}.

\section{\label{sec:conclusion}Conclusion}

Generative artificial intelligence tools such as LLMs are fundamentally ill-equipped to serve directly as arbiters of truth, because they sample from a probability distribution over the data used to train them, i.e., received and mostly conventional wisdom that is permeated with bias. 
However, these tools can serve ``fast'' or ``system 1'' reasoning purposes while more classically algorithmic techniques perform ``slow'' or ``system 2'' reasoning \cite{kahneman2011thinking}. While CDI (or any other near-term approaches, particularly those based on learning \cite{van2024reclaiming}) will not lead immediately to artificial general intelligence, CDI does provide a good model for many forms of cognition, including perception and planning \cite{blokpoel2025theoretical}. 

Unusually among artificial intelligence techniques, CDI is
\begin{itemize}
    \item explainable: cuts are readily interpretable; near-optimal cuts can be compared.
    \item ethical: ethical guidelines can be stipulated propositions \cite{sivaraj2019cogent}.
    \item reproducible and stable: medians of LLM-generated coherence graphs converge nicely (at least for small scales). To the extent that substantially different graphs are obtained, they are presumably \emph{cut sparsifications} of a denser implicit ``Platonic'' structure \cite{benczur1996approximating, spielman2011spectral,soma2019spectral}.
    \item versatile: any sufficiently capable multimodal model can be employed.
    \item capable of handling abstraction: CDI with LLMs operates over text (or multimodal) representations.
    \item capable of handling ambiguity: CDI amounts to \emph{resolving} ambiguity.    
    \item based on a flexible cognitive model grounded in decades of research: there are deep connections with practical psychology, law, philosophy of science, etc. \cite{holyoak1999bidirectional,joseph2009coherence,criado2016coherence}.
    \item generalizable by construction: mathematical considerations led to an independent, general formulation \cite{huntsman2025neurosymbolic}.
\end{itemize}
CDI also offers a plausible approach for automatically making sense of competing arguments in a way that accords with the features enumerated here.

This paper is part of an argument that it is now feasible to computationally instantiate a reasonable approximation of a coherence theory of truth \cite{sep-truth-coherence}: the recent benchmark \cite{huntsman2025neurosymbolic} provides additional quantitative evidence in this direction. By ``hard-coding'' acceptance of conclusively established propositions, this theory can furthermore be anchored in a correspondence theory of truth \cite{sep-truth-correspondence}. In other words, coherence computations can be required to incorporate privileged information that also coheres with observed reality. While it is easy to imagine attempts to try the same thing with privileged information that does not cohere with observed reality, lies cannot persist when they can easily be unraveled.

Even with flawless technology (which this will not be), obstacles will be manifold. For example, in a pluralistic society, legal coherence may actually require sacrificing fairness in some ways \cite{perez2005institutionalization}. Ultimately, people must decide matters for themselves. It is only reasonable to hope that technology can serve as a reliable tool to help people make their decisions more coherent.








\appendices

\section{\label{sec:l1_distances} $L^1$/graph edit distance convergence}

\begin{figure}[htbp]
    \centering
    \includegraphics[width=1\linewidth, trim={0 0 0 0mm}, clip]{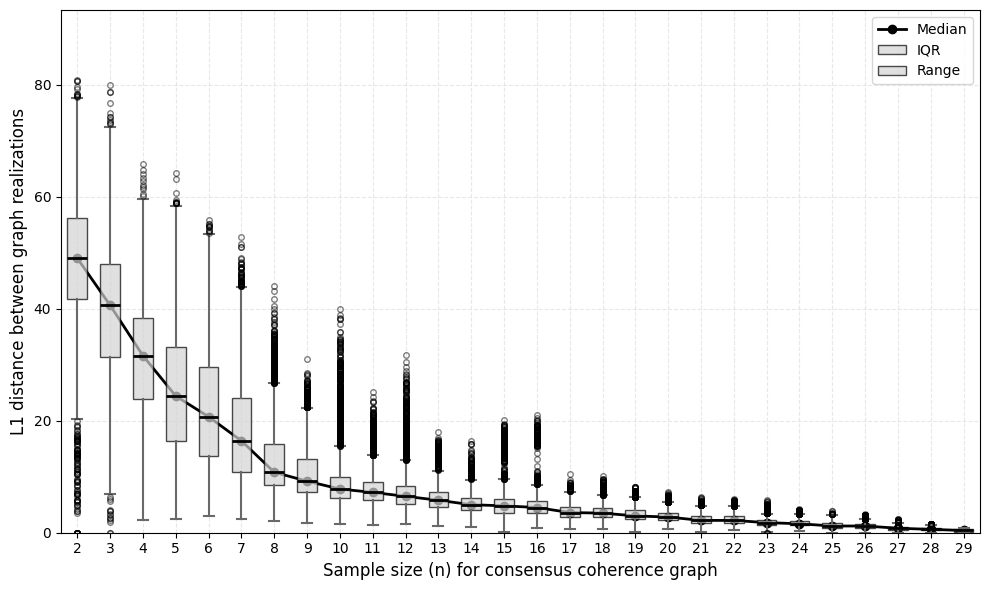}
        \caption{$L^1$/graph edit distance between medians of $n$ of $N = 30$ coherence graphs used to construct the consensus of $N$ coherence graphs in Figure \ref{fig:melianDialogue}. This sort of rapid convergence is typical but not ubiquitous.}
    \label{fig:l1_distances}
\end{figure}

Figure \ref{fig:l1_distances} shows how taking a median of multiple coherence graphs can yield a stable, broadly reproducible result. Starting from $N = 30$ coherence graphs, for each $1 < n < N$ we consider $\min(100,\binom{N}{n})$ subsamples of size $n$ and compute their distribution of distances from the median of $N$. The median of $n$ trivially converges to the median of all $N$ graphs, but if the convergence is rapid, then we can have confidence in the stability and reproducibility of the median of $N$. On the other hand, if convergence is slow until near $N$, then the result may not be robust: this can be addressed by, e.g., increasing $N$ and trying again, seeking to reduce the number of propositions, or some other tactic as appropriate. 

We recall that taking medians is optimal for $L^1$ distance, which in this context also equals the graph edit distance.

\section{\label{sec:prompt_practical} Prompt for practical examples}

The prompt we used to build coherence graphs from a formatted set of propositions is in Table \ref{tab:prompt_practical}. Its verbosity owes to descent from prompts used to gauge models' ability to evaluate pairwise consistency of propositions without reference to external facts \cite{huntsman2024prospects}.

\begin{table}[htbp]
\caption{Prompt for obtaining individual coherence graphs}
\begin{center}
\begin{tabular}{|l|}
\hline
Imagine that you are a perfectly objective arbitrator with impeccable \\
judgment and integrity. In response to a prompt of the form \\ 
'buildCoherence: ' below followed by a list of labeled propositions, \\
please do the following: First, determine which pairs of propositions \\
are substantively related. Second, for each related pair of propositions, \\
determine their logical relationship, assuming that at least one is true, \\
whether or not either actually is. I want you to ignore the truth, falsity \\
or basis in fact of either claim. Third, based on your determination just \\
above, numerically rate the relative consistency of the two propositions. \\
Do not pay attention to or comment on the truth or basis in fact of \\
either proposition independent of the other. Your rating of relative \\
consistency should be on a scale from 0 to 10, with a value of 0 for a \\
pair of propositions that are not at all consistent and a value of 10 for \\
a pair of propositions that are totally consistent. I cannot emphasize \\
enough that for your rating, I want you to ignore the truth or basis in \\
fact of either proposition, since anything that is not consistent with \\
reality cannot be true. If you determine that propositions are unrelated \\
despite previously determining otherwise, omit that pair. To be clear, a \\
pair of false but consistent claims should also be rated a 10. Meanwhile, \\
a pair of propositions of which one is true and the other is false, should \\
be rated a 0. Finally, construct a NetworkX graph where propositions \\
are vertices and edges correspond to substantively related pairs of \\
propositions, with weights given by the consistency ratings just above. \\
Only return the edge list with proposition labels for vertices. i.e., \\
return responses in this format (here 'p2', 'p3', 'p4', and 'p5' are \\
labels): \\
$[$('p2', 'p3', 0), ('p2', 'p5', 10), ('p3', 'p4', 9), ('p3', 'p5', 2)$]$. \\
Order vertices (in edges) and edges (in the graph) lexicographically.\\ 
\\ 
buildCoherence: \\
\hline
\end{tabular}
\label{tab:prompt_practical}
\end{center}
\end{table}

\section*{Acknowledgments}
Thanks to Ludmilla Huntsman, Michael Robinson, Jewell Thomas, and Benjamin Wittes; and also to an exceptionally careful reviewer. The original results and views expressed herein are the author's own and do not constitute endorsement by any other party.


\bibliography{arguments}
\bibliographystyle{IEEEtran}

\end{document}